\documentclass{IEEEtran}
\usepackage{silence}
\usepackage{amsmath,amssymb,amsfonts}
\usepackage{amsthm}
\usepackage{algorithmic}
\usepackage{algorithm}
\usepackage{graphicx}
\usepackage{graphbox}
\usepackage{textcomp}
\usepackage{hyperref}
\usepackage{multirow}
\usepackage{balance}
\usepackage{xcolor}
\usepackage[export]{adjustbox}
\usepackage{color}
\usepackage{accents}
\usepackage{cite}
\usepackage{dsfont}
\usepackage{textcomp}
\usepackage{comment}
\usepackage{bm}
\usepackage{upgreek}
\usepackage{url}

\usepackage[shortlabels]{enumitem}

\def\Ub{{\mathbf{R}}}
\def\RR{{\mathbb R}}

\def\NN{\mathbb{N}}

\def\sign{\,\mathrm{sign}}

\newcommand{\abs}[1]{\left|#1\right|}

\newcommand{\sbrm}[1]{\sb{\mathrm{#1}}}

\newcommand{\changed}[1]{#1}

\newcommand{\gammamax}{\overline\gamma}
\newcommand{\Nmax}{\overline N}
\newcommand{\kmax}{\overline k}
\newcommand{\Tmax}{\overline T}

\definecolor{revXIII}{HTML}{D45500}
\newcommand{\Rev}[1]{\textcolor{black}{#1}}
\newcommand{\Revv}[1]{\textcolor{black}{#1}}

\newtheorem{theorem}{Theorem}

\newtheorem{lemma}{Lemma}
\newtheorem{remark}{Remark}
\newtheorem{definition}{Definition}
\newtheorem{proposition}{Proposition}

\newcommand{\F}{\mathcal{F}}
\newcommand{\FL}{\F_L}
\newcommand{\EN}{\mathcal{E}_N}
\newcommand{\meas}{u}

\newcommand{\Diff}{\mathcal{D}}
\newcommand{\Tc}{\mathcal{T}}

\newcommand{\yw}{y\sbrm{w}}
\newcommand{\Diffw}{\Diff\sbrm{w}}
\newcommand{\ym}{y\sbrm{m}}
\newcommand{\Diffm}{\Diff\sbrm{m}}
\newcommand{\etam}{\eta\sbrm{m}}

\def\BibTeX{{\rm B\kern-.05em{\sc i\kern-.025em b}\kern-.08em
    T\kern-.1667em\lower.7ex\hbox{E}\kern-.125emX}}
\begin{document}

\title{Optimal robust exact first-order differentiators with Lipschitz-continuous output
\thanks{Corresponding author: Richard Seeber (richard.seeber@tugraz.at)}
}
\author{Rodrigo Aldana-L\'opez, Richard Seeber, Hernan Haimovich, and David G\'omez-Guti\'errez%
\thanks{\textcolor{red}{This is the accepted version of the manuscript: ``Optimal robust exact first-order differentiators with Lipschitz-continuous output,"R. Aldana-López, R. Seeber, H. Haimovich and D. Gómez-Gutiérrez, in IEEE Transactions on Automatic Control, 2023, DOI: 10.1109/TAC.2025.3555481. 
\textbf{Please cite the publisher's version}. For the publisher's version and full citation details see:
\href{https://doi.org/10.1109/TAC.2025.3555481}{https://doi.org/10.1109/TAC.2025.3555481}. 
}}
\thanks{*Work supported by Agencia I+D+i, Argentina, under grant PICT 2021-I-A-00730, and by Christian Doppler Research Association, Austrian Federal Ministry for Digital and Economic Affairs, and National Foundation for Research, Technology and
Development.}%

\thanks{R. Aldana-L\'opez and D. G\'omez-Guti\'errez are with Intel Tecnolog\'ia de M\'exico, Intel Labs, Intelligent Systems Research Lab, Jalisco, Mexico. D.~G\'omez-Guti\'errez is also with Tecnológico Nacional de México, Instituto Tecnológico José Mario Molina Pasquel y Henríquez, U. A. Zapopan, Jalisco, Mexico. 
        (e-mail: rodrigo.aldana.lopez@gmail.com, david.gomez.g@ieee.org)}%
\thanks{R. Seeber is with the Christian Doppler Laboratory for Model Based Control of Complex Test Bed Systems, Institute of Automation and Control, Graz University of Technology, Graz, Austria.
        (e-mail: richard.seeber@tugraz.at)}%
\thanks{H. Haimovich is with Centro Internacional Franco-Argentino de 
  Ciencias de la Informaci\'on y de Sistemas (CIFASIS)
  CONICET-UNR, 2000 Rosario, Argentina.
        (e-mail: haimovich@cifasis-conicet.gov.ar)}%
}

\maketitle

\begin{abstract}
The signal differentiation problem involves the development of algorithms that allow to recover a signal's derivatives from noisy measurements. This paper develops a first-order differentiator with robustness to measurement noise, exactness in the absence of noise, optimal worst-case differentiation error, and Lipschitz-continuous output where the output's Lipschitz constant is a tunable parameter. This combination of advantageous properties is not shared by any existing differentiator. \Rev{A sample-based implementation by implicit discretization is obtained, which is quasi-exact, inherits the optimal worst-case error bound, and the Lipschitz constant translates to a discrete-time increment bound. Illustrative examples are} provided to highlight the features of the developed differentiator. \Rev{Open-source code for our differentiator can be found in \url{https://github.com/RodrigoAldana/OREdiff}.}

\end{abstract}

\section{Introduction}

The online estimation of a signal's derivative from noisy measurements is a fundamental problem in control theory for its application in fault diagnosis, identification, observation~\cite{Fraguela2012}, and control. Different strategies, typically known as \emph{differentiators}, have been proposed based on the information about the signal and the noise. In this paper, we are interested in the case where the only available information is that the second-order derivative of the signal has a known bound $L$ and that an unknown constant $N$ bounds the noise signal. For this scenario, different approaches have been proposed in the literature, such as high-gain linear algorithms~\cite{Vasiljevic2006DifferentiationNoise}, algorithms based on unbounded Time-Varying Gains (TVGs)~\cite{Holloway2019,Orlov2022Prescribed-TimeGains}, and algorithms based on sliding-mode control~\cite{Levant1998RobustTechnique,cruz2011uniform,seeber2021robust}, such as the so-called super-twisting algorithm~\cite{Levant1998RobustTechnique}. 

Qualitative properties and quantitative metrics can assess the performance of a differentiator to evaluate its suitability for a specific application. 
Two important qualitative properties are \emph{exactness} and \emph{robustness}~\cite{Levant1998RobustTechnique,seeber2023}. A differentiator is exact if, in the absence of noise, its output converges in finite-time to the actual signal's derivative; robust if the behavior in the presence of noise uniformly converges to the behavior in the absence of noise, as the noise magnitude tends to zero.

A related essential feature is the class of convergence to the actual signal's derivative in the noise-free case, which can be asymptotic, in a finite time, in a fixed time (a finite-time uniformly bounded for every initial condition), and from the beginning (except at the initial time). Moreover, an essential quantitative metric is the differentiator's \emph{worst-case accuracy}, which, in general, cannot be better than $2\sqrt{NL}$ and in the case of an exact differentiator, cannot be better than $2\sqrt{2NL}$~\cite{seeber2023}.

Differentiators with unbounded TVGs are exact, with a fixed-time convergence prescribed by the user~\cite{Holloway2019,Orlov2022Prescribed-TimeGains}, but they are not robust and exhibit unbounded worst-case accuracy~\cite{Aldana-Lopez2022OnAlgorithms}. The super-twisting algorithm~\cite{Levant1998RobustTechnique} is robust and exact and features finite-time convergence that grows unbounded as a function of the initial condition, and worst-case accuracy proportional to $\sqrt{NL}$. However, it cannot reach the optimal worst-case accuracy $2\sqrt{2NL}$ and, moreover, tuning it to improve its worst-case accuracy reduces its convergence speed~\cite{seeber2023worst}. 
Linear high-gain differentiators are robust but not exact, converging to the actual signal's derivative only when the bound on its second-order derivative is $L=0$ and only asymptotically. The algorithms in~\cite{cruz2011uniform,moreno2021arbitrary} are robust, exact, and fixed-time convergent; their approach is based on homogeneity in the bi-limit, and achieve the worst-case accuracy of the super-twisting differentiator but for ``small signals''. Finally, the algorithm proposed in~\cite{seeber2021robust} is robust, exact, and fixed-time convergent, but its worst-case accuracy has yet to be analyzed.

The only exact differentiator known to achieve the optimal worst-case accuracy $2\sqrt{2NL}$ was proposed recently in~\cite{seeber2023}; this differentiator is exact from the beginning and robust almost from the beginning. The approach is based on a single-parameter adaptation of a finite-difference differentiator~\cite{Diop2000numerical,Levant2007finite}. However, in contrast to the Lipschitz-continuous output of the super-twisting algorithm~\cite{Levant1998RobustTechnique} and to the smooth output in~\cite{Vasiljevic2006DifferentiationNoise}, the output of the differentiator in~\cite{seeber2023} features a direct feed-through from the noise that causes the estimate to inherit the noise's discontinuous nature. When the derivative estimate is used for control, the latter feature may be detrimental by inducing high-frequency vibrations in the actuator~\cite{Levant2010chattering,Utkin2015discussion}. Moreover, this differentiator lacks robustness at the initial time instant, which may lead to an unbounded initial output signal.

Implementing a differentiator on a digital computer requires developing algorithms based on sampling the signal of interest. \Rev{One approach is discretizing a continuous-time algorithm, as in~\cite{brogliato2021digital,carvajal2021implicit,hanan2021low}, which is a challenging problem for sliding-mode algorithms. Implicit (backward) Euler discretization results in better accuracy and lower numerical chattering than other explicit methods as compared in~\cite{brogliato2021digital}}.

This paper develops a robust and exact differentiator that has a Lipschitz-continuous output and achieves the optimal worst-case accuracy, thus overcoming the limitations of~\cite{seeber2023}. This combination of features is not shared by any existing differentiator. %
The proposed differentiator is obtained by combining the differentiator in~\cite{seeber2023} with a first-order sliding-mode filter to obtain robustness and a Lipschitz-continuous output while retaining the optimal accuracy of~\cite{seeber2023}. \Rev{From the continuous-time design we then obtain a sample-based implementation by implicit discretization, which retains the continuous-time features to the largest possible extent. In particular, the sample-based differentiator is quasi-exact, cf. \cite[Definition 6.4]{seeber2023}, inherits the optimal worst-case error bound, and the Lipschitz constant translates to a discrete-time increment bound. %
}
The proposed differentiator's tuning is very simple: in addition to the knowledge of the signal's second-order derivative bound $L$ \Rev{and a rough estimate $\Nmax$ of the noise bound $N$}, an additional single parameter $\gamma>L$ has to be selected, limiting the differentiator's output's Lipschitz constant and hence trades (fixed-time) convergence speed for output ``smoothness'', but does not otherwise impact the differentiation accuracy. 
The main features of the developed differentiator, namely fixed-time convergence, robustness and optimal worst-case accuracy, are theoretically established. 

Compared to the conference version~\cite{aldana2023exact}, this paper formally analyses the convergence time and robustness of the proposed differentiator and \Rev{includes the} novel sample-based implementation. The worst-case accuracy and the convergence speed of the resulting discrete algorithm are analyzed. Moreover, we introduce a modification that yields a uniformly bounded, with respect to the initial conditions, convergence time bound. 

The rest of the paper is organized as follows: Section~\ref{sec:prel} introduces the problem and recalls the differentiator proposed in~\cite{seeber2023}. Based on such a differentiator, Section~\ref{sec:continuousDiff} proposes a continuous-time differentiator with Lipschitz-continuous output, whereas a sample-based implementation is given in Section~\ref{sec:discreteDiff}. Numerical comparison against state-of-the-art algorithms are given in Section~\ref{sec:simulations}. Finally, Section~\ref{sec:conclusions} presents the conclusions. \Rev{All proofs are collected in the Appendix.}

\textbf{Notation:}
$\RR_{>0}$, $\RR_{\ge 0}$ and $\RR$ denote the postive, the nonnegative, and the whole real numbers, respectively; $\NN$ denotes the natural numbers.
$\lceil a \rceil$ denotes the least integer not less than $a\in\RR$. One-sided limits of a function $f$ at time instant $T$ from above are written as $\lim_{t \to T^+} f(t)$, $\limsup_{t \to T^+} f(t)$, and $\liminf_{t \to T^+} f(t)$. If $\alpha\in\RR$, then $|\alpha|$ denotes its absolute value. ‘Almost everywhere’ is abbreviated as ‘a.e.’. %

\section{Preliminaries}
\label{sec:prel}
In this section, we introduce the performance-related definitions of \emph{worst-case error}, \emph{exactness}, and \emph{accuracy}, which are recalled from \cite{seeber2023} for the most part. Moreover, the differentiator in~\cite{seeber2023} is presented as a basis for subsequent developments. Afterwards, the problem statement to be addressed in this work is introduced.

\subsection{Performance Measures for Differentiators}

Denote with $\F$ the set of differentiable functions $f : \RR_{\ge 0} \to \RR$ with Lipschitz-continuous derivative $\dot f$ on all $\RR_{\ge 0}$. We are interested in the differentiation of functions $f\in\mathcal{F}$ using noisy measurements $u = f+ \eta$ where $\eta$ is a uniformly bounded noise signal. Henceforth, the classes of signals to consider, from which measurements are generated, are \Revv{given by}
\begin{subequations}
\label{eq:defFE}
    \begin{align}
        \FL = \{ f \in \F &: \abs{\ddot f(t)} \le L \text{ a.e. on } \RR_{\ge 0}\}\Revv{,} \\
        \EN = \{ \eta \in \mathcal{E} &: \abs{\eta(t)} \le N \text{ for all }t\ge 0 \}.
    \end{align}
\end{subequations}
where $\mathcal{E}$ denotes the set of all uniformly bounded functions $\eta : \RR_{\ge 0} \to \RR$ on $\RR_{\ge 0}$. \Rev{Note that noises $\eta\in\mathcal{E}_N$ are not required to be Lebesgue measurable.}
Write $\FL + \EN = \{ f + \eta : f \in \FL, \eta \in \EN\}$ for the set of inputs $u$ with fixed $L$ and $N$.
Hence, the following set contains all possible inputs to be considered for the differentiator:
\begin{equation}
\label{eq:U}
    \mathcal U = \bigcup_{\substack{L \ge 0 \\ N \ge 0}} (\FL + \EN).
\end{equation}

A differentiator is a causal operator,  \Rev{cf.\cite[Definition 2.1]{seeber2023}},  $\Diff : \mathcal U \to ( \RR_{\ge 0} \to \RR)$ mapping a signal $\meas\in\mathcal{U}$ to an estimate $\Diff \meas$ for the derivative of $f$. 
For future reference, for every $R\ge 0$, define the class of signals with a bounded second derivative that, in addition, have a bounded initial value and initial derivative%
\Rev{
\begin{equation}
\label{eq:flr}
    \FL^\Ub := \{ f \in \FL : |f(0)| \le R_0, |\dot f(0)| \le R_1 \}\Revv{,}
\end{equation}
with $\Ub=[R_0,R_1]$.} The next definitions recall concepts %
that are useful to describe the features required for a differentiator in this work.

\begin{definition}[Worst-case error \cite{seeber2023}]
\label{def:worst-case}
Let $L, N\in \RR_{\ge 0}, \Ub\in\RR_{\ge 0}^2$. A differentiator $\Diff$ is said to have worst-case error $M_N^{L,\Ub}(t)$ from time $t\ge 0$ over the signal class $\FL^\Ub$ with noise bound $N$ if
        \begin{equation}
                M_N^{L,\Ub}(t) =\sup_{\substack{f\in\FL^\Ub \\ \eta\in\EN}} \sup_{\tau \ge t} \big|\dot f(\tau) - [\Diff (f+\eta)](\tau)\big|.
        \end{equation}
\end{definition}

\begin{definition}[Exactness \cite{seeber2023}]\label{def:exactness}
A differentiator $\mathcal{D}$ is said to be exact in finite time over $\FL$, if for each $\Ub \in \RR_{\ge 0}^2$ there exists %
$\hat{\mathcal{T}}(\Ub) \in \RR_{>0}$ such that $M_0^{L,\Ub}(\hat{\mathcal{T}}(\Ub)) = 0$. The differentiator is said to be exact in fixed time, if there exists $\hat{\mathcal{T}}\in\mathbb{R}_{>0}$ such that $M_0^{L,\Ub}(\hat{\mathcal{T}}) = 0$ for all $\Ub\in\mathbb{R}_{\geq 0}^2$.
\end{definition}
\begin{definition}[Robustness \cite{seeber2023}]
A differentiator $\mathcal{D}$ is said to be  robust from the beginning over $\mathcal{F}_L$ if, for all $\Ub\in\mathbb{R}_{\geq 0}^2$,
\begin{equation}
\label{eq:robustQ}
\begin{aligned}
    &Q_N^{L,\Ub}(t) = \sup_{\substack{f\in\FL^\Ub \\ \eta\in\EN}} \sup_{\tau \ge t} \big|[\mathcal{D}f](\tau) - [\mathcal{D} (f + \eta)](\tau)\big| \\
    &Q^{L,\Ub}(t) = \limsup_{N \to 0^+} Q_N^{L,\Ub}(t)
\end{aligned}
\end{equation}
fulfills $Q^{L,\Ub}(0) = 0$. In addition, $\mathcal{D}$  is robust almost from the beginning if $Q^{L,\Ub}(t) = 0$ for all $\Ub\in\mathbb{R}_{\geq 0}^2, t>0$.
\end{definition}

The time $\hat{\mathcal{T}}$ in Definition~\ref{def:exactness} is called a convergence time bound of the differentiator and relates to the case without measurement noise.
In the following, the notion of convergence-time functions \emph{in the presence of noise} is introduced, based on bounds for the asymptotic accuracy $C_L$ as defined in \cite[Definition~3.6]{seeber2023}.
Loosely speaking, such a function bounds from above the time after which the differentiator with noisy input achieves the corresponding accuracy.

\begin{definition}[Accuracy]
    \label{def:accuracy}
    A differentiator $\Diff$ is said to have 
        accuracy bound $\hat{C}_{L,\Nmax} \in \RR_{\ge 0}$ for signals in $\FL$ with noise bounds less than $\Nmax \in \RR_{> 0}$, if there exists a function $\hat{\Tc} : \RR_{\ge 0}^2 \times [0, \Nmax) \to \RR_{\ge 0}$ continuous in its second argument where
    \begin{equation}
        \label{eq:rel-ac-bnd-single}
        M_N^{L,\Ub}[\hat{\Tc}(\Ub,N)] \le \hat{C}_{L,\Nmax}  \sqrt{N L},
    \end{equation}
    holds for all $N \in [0, \Nmax)$ and $\Ub\in\mathbb{R}_{\geq 0}^2$.
    In this case, $\hat{\Tc}$ is called a convergence time function in the presence of noise for $\hat{C}_{L,\Nmax}$.
\end{definition}
\begin{remark}
    Note that any accuracy bound as defined above is an upper bound for the asymptotic accuracy $C_L$ defined in \cite[Definition~3.6]{seeber2023}, i.e., $C_L \le \hat C_{L,\Nmax}$.
    Hence, according to \cite[Proposition~3.10]{seeber2023}, the lowest possible (i.e., optimal) value of $\hat C_{L,\Nmax}$ is given by $2 \sqrt{2}$.
\end{remark}

\subsection{Optimal exact differentiation}
In this section, we recall the differentiator from \cite{seeber2023} with parameter $\gammamax = 1$, which has the advantageous features of being optimal and exact. This differentiator, denoted here by $\Diffw$ with output $\yw=\Diffw u$, is given by
\begin{subequations}
    \label{eq:proposed:diff}
    \begin{align}
    \label{eq:proposed:diff:y}
        &\yw(t) = \begin{cases}
            0 & \text{if } t = 0 \\
            \lim_{T \to 0^+} \dfrac{u(t) - u(t - T)}{T} & \text{if } t > 0, \hat T(t) = 0 \\
            \dfrac{u(t) - u(t - \hat T(t))}{\hat T(t)} & \text{if } t > 0, \hat T(t) > 0,
        \end{cases} \\
        \intertext{with time difference $\hat T(t)$ adapted according to}
                \label{eq:def:That:t}
        &\hat T(t) =  \min\biggl\{t, \Tmax, 2  \sqrt{ \frac{\hat N(t)}{L}}\biggr\},
        \displaybreak[0]
        \intertext{where $\hat N(t)$ is an estimate for the noise amplitude that is determined from the measurement $u$ according to}
        \label{eq:def:Nhat:t}
&\hat N(t) = \frac{1}{2} \sup_{\substack{T \in (0, \Tmax] \\ T \le t \\ \sigma \in [0, T]}} \biggl( \abs{Q(t, T, \sigma)} - \frac{L \sigma (T - \sigma)}{2} \biggr),\\
        \intertext{with $Q(t, T, \sigma)$ defined as}
        \label{eq:def:aT:t}
        &Q(t, T, \sigma) = u(t - \sigma) - u(t) + \frac{u(t) - u(t-T)}{T} \sigma.  
\end{align}
        
The differentiator is characterized by the window-length parameter $\Tmax\in\mathbb{R}_{\geq 0}$ indicating the extent of $u$'s historical data used in output calculation. \Rev{The form of the estimate $\hat{N}(t)$ is explained in detail in \cite[Section 5.3]{seeber2023}, noting that the noise amplitude can be estimated by checking all measurement deviations in \eqref{eq:def:aT:t} by sweeping over $T,\sigma$.}
        
\end{subequations}

In \cite{seeber2023}, it is proven that $\yw(t)$ is well-defined for every $t \geq 0$ and for any $u\in\mathcal{U}$. This also ensures that the limit in \eqref{eq:proposed:diff:y} exists. Additionally, this differentiator is exact from the beginning\footnote{For the formal definition of exactness from the beginning, refer to \cite[Definition~2.4]{seeber2023}.} and achieves the optimal asymptotic accuracy bound $\hat{C}_{L,\Nmax} = 2\sqrt{2}$ for all $\Nmax \in [0,L \Tmax^2/2)$, with the convergence time function $\hat{\mathcal{T}}(\Ub,N) = \sqrt{2N/L}$. Nonetheless, the output of this differentiator may lack continuity under certain noise features. The reason for this is that the noise $\eta$, which may be discontinuous, is directly incorporated into the formula for $\yw(t)$ in \eqref{eq:proposed:diff:y} via the input term $u = f + \eta$.

The following result provides an additional auxiliary bound on the differentiation error $\dot f - \Diffw u$, which is important for subsequent developments in this work.

\begin{lemma}
\label{lem:error:boundedness}
    Let $L \in \RR_{> 0}$ and consider the differentiator $\Diffw$ defined in \eqref{eq:proposed:diff} with parameter $\bar T \in \RR_{>0} \cup \{ \infty \}$.
    Then, for all $u \in \mathcal{U}$ with noise bounds less than $\overline{N}=\frac{L\overline{T}^2}{2}$ and all $t > 0$, the differentiator output $\yw = \Diffw u$ satisfies
    \begin{equation}
    \label{eq:boundsyw}
        |\yw(t)-\dot{f}(t)| \le \begin{cases}
        2\sqrt{2NL} & t\geq \sqrt{2N/L}\\
                Lt + \frac{2N}{t} & \text{otherwise}. \\
        \end{cases}
    \end{equation}
\end{lemma}
The proofs for all lemmata are given in the appendix.

\subsection{Problem statement}

 The problem addressed in this paper is the following. Let $L > 0$ be known \Rev{and assume a rough estimate $\Nmax$ for the bound of the noise magnitude}. Design a differentiator $\mathcal{D}$ with the following features:
\begin{enumerate}[i)]
    \item $\mathcal{D}$ has \emph{Lipschitz-continuous output}\label{item:prob:Lipschitz};
    \item $\mathcal{D}$ is robust from the beginning;\label{item:prob:robust}
    \item $\mathcal{D}$ has optimal accuracy bound $\hat C_{L,\Nmax} = 2 \sqrt{2}$ with an $\Nmax$ that can be made arbitrarily large by appropriate tuning;\label{item:prob:optimal}
    \item $\mathcal{D}$ is exact in fixed time over $\FL$;\label{item:prob:exact}
\end{enumerate}
and provide a sample-based implementation of that differentiator that retains the above features to the largest extent possible.

Existing exact sliding-mode differentiators, such as the super-twisting differentiator \cite{Levant1998RobustTechnique} and its variants, do not fulfill item \ref{item:prob:optimal} as shown in \cite[Proposition 3.1]{seeber2023worst}, whereas the optimal exact differentiator from \cite{seeber2023} does not fulfill item~\ref{item:prob:Lipschitz}. 

In this work, we present a differentiator complying with all the previous features. A similar problem statement was studied in \cite{aldana2023exact}, with the difference that here, we are interested in robustness and fixed-time convergence instead of finite-time. In addition, different to \cite{aldana2023exact}, we discuss a discrete-time implementation for the proposed differentiator, with analogous features to those exhibited in continuous-time.

\section{\Rev{Continuous-time design}}
\label{sec:continuousDiff}

Upon comparing the characteristics of the differentiator $\Diffw$ as outlined in \eqref{eq:proposed:diff} with the requirements specified in our problem statement, it becomes evident that the goal is to balance a decrease in the speed of achieving exactness (opting for finite time rather than immediate exactness) against the advantage of having a Lipschitz-continuous output. The fundamental strategy involves filtering the output $\yw = \Diffw u$ using a first-order sliding-mode system. To facilitate this, a modified version of the differentiator, $\Diffm$, is introduced as a regularization of $\Diffw$ in Section~\ref{sec:regularization}. Following this, the proposed differentiator $\Diff$, designed to produce a Lipschitz-continuous output, is detailed in Section~\ref{sec:main}.

\subsection{Differentiator output regularization}
\label{sec:regularization}

The output of the differentiator $\Diffw$ of~\eqref{eq:proposed:diff}, namely $\yw$ in~\eqref{eq:proposed:diff:y}, may lack not only continuity but also (Lebesgue) measurability. To ensure that a filter that takes $\yw$ as input has a well-defined solution, $\yw$ must be (at least) a measurable function. One of the main reasons for this lack of measurability is the fact that the supremum in~\eqref{eq:def:Nhat:t} is taken over an uncountable set, \Rev{apart from the fact that the noise $\eta$ is not assumed to be Lebesgue measurable in the present paper}. 
To ensure measurability, we introduce the following regularization. For any function $v:\mathbb{R}_{\geq 0}\to\mathbb{R}$, its regularization $v^\ddag : \mathbb{R}_{\ge 0} \to \mathbb{R} \cup \{-\infty,\infty\}$ is defined as %
\begin{equation}
\label{eq:measDiff:operator}
    v^\ddag(t) = \begin{cases}
    v(0) & \text{if $t = 0$} \\
    \frac{\limsup_{\varepsilon\to 0^+}{v}(t-\varepsilon) + \liminf_{\varepsilon\to 0^+}{v}(t-\varepsilon)}{2} & \text{if $t > 0$},
    \end{cases}
\end{equation}
with $\infty + (-\infty) := 0$ in case both limits are infinite. It was proven in \cite[Lemma 2]{aldana2023exact} that if $v : \RR_{\ge 0} \to \RR$ is locally bounded on $\RR_{> 0}$, then $v^{\ddag}$ in \eqref{eq:measDiff:operator} takes only finite values, is locally bounded on $\RR_{> 0}$, and is Lebesgue measurable.

Define a new, intermediate differentiator $\Diffm$ whose output $\ym = \Diffm u$ is a regularized version of $\yw$, namely 
\begin{equation}
\label{eq:measDiff}
    \ym(t) = \yw^\ddag(t).
\end{equation}
The following result is an extension of \cite[Proposition 1]{aldana2023exact}, ensuring some advantageous features of $\Diffm$.

\begin{proposition}
\label{prop:proposed:exact}
Let $L \in \RR_{>0}$ and consider the differentiator $\Diffm$ with output $\ym = \Diffm u$ defined by \eqref{eq:proposed:diff} and \eqref{eq:measDiff}, with parameter $\Tmax \in \RR_{> 0} \cup \{ \infty \}$. Then, \changed{the following statements are true:}
\begin{enumerate}[a)]
    \item the output $\Diffm u$ is Lebesgue measurable for all $u\in\mathcal{U}$;\label{item:meas}
    \item %
    the worst-case differentiation error of $\Diffm$ fulfills 
    $$
    M_N^{L,\Ub}(t) \le \begin{cases}
        Lt + \frac{2N}{t} & \text{if }t\in(0,\sqrt{2N/L}) \\
        2\sqrt{2NL} & t\geq \sqrt{2N/L},
        \end{cases}
    $$ for all $N \in [0, L \Tmax^2/2)$.\label{item:errorbound} %
    \item\label{item:robustDw} The differentiator $\Diffm$ is robust almost from the beginning.
    \end{enumerate}
\end{proposition}

\subsection{Exact differentiator with Lipschitz-continuous output}
\label{sec:main}

Define the output of the proposed differentiator as the Filippov \cite{Filippov1988} solution to
\begin{subequations}
    \label{eq:aux:dyn:sys}
\begin{equation}
\dot{y}(t) = -\gamma \sign(y(t)-\ym(t)),\quad
y(t_0)=\ym(t_0),
\end{equation}
and as
\begin{equation}
    y(t) = 0 \qquad \text{for $t \in [0, t_0)$},
\end{equation}
\end{subequations}
with two design parameters $\gamma>0$ and $t_0 \ge 0$, i.e., by applying a first-order sliding-mode filter to the output $\ym$ of the regularization.
Note that $\ym$ may be unbounded in a right-neighborhood of $t = 0$; nevertheless, the right-hand side of \eqref{eq:aux:dyn:sys} is uniformly bounded by virtue of the sign function and Lebesgue measurable as a consequence of $\ym$ being Lebesgue measurable according to Proposition~\ref{prop:proposed:exact}-\ref{item:meas}. \Rev{Moreover, as discussed formally in the following, the use of first-order sliding-modes in \eqref{eq:aux:dyn:sys} ensures the Lipschitz-continuity of the output. }
The proposed differentiator is then defined by \eqref{eq:proposed:diff}, \eqref{eq:measDiff}, and \eqref{eq:aux:dyn:sys}.

The following establishes some properties of the proposed differentiator, which, different to \cite{aldana2023exact}, is extended to allow an arbitrary initialization time $t_0\geq 0$ in \eqref{eq:aux:dyn:sys} that may be different to that used in the previous stage $\mathcal{D}_\mathrm{m}$. In addition, it ensures that the proposed differentiation is robust from the beginning.

\begin{theorem}
\label{thm:proposed:filtered}
Let $L >0, N\geq 0$ and consider the differentiator $\Diff$ with output $y = \Diff u$ defined by \eqref{eq:proposed:diff}, \eqref{eq:measDiff}, and \eqref{eq:aux:dyn:sys} with parameters $\Tmax \in \RR_{> 0} \cup \{ \infty \}$, and $\gamma>L$. Then, \changed{the following statements are true:}
\begin{enumerate}[a)]
    \item\label{item:filt} the output of $\Diff$  is Lipschitz-continuous on $[t_0, \infty)$ as well as on $[0, t_0)$  for any $u\in\mathcal{U}$.
    \item\label{item:errorbound:filt}%
    if $t_0 = 0$, then $\Diff$ is exact in finite time and has accuracy bound $\hat C_{L,\Nmax} = 2 \sqrt{2}$ for signals in $\FL$ with noise bounds less than $\Nmax = \frac{L \Tmax^2}{2}$, with corresponding convergence time function in the presence of noise given by
    \begin{equation}
    \label{eq:setting:time:t0}
    \hat{\mathcal{T}}(\Ub,N) = 2 \sqrt{\frac{2N}{L}} + \frac{R_1}{\gamma - L}.
    \end{equation}
    \item \label{item:settling} if $t_0 > 0$, then $\Diff$ is exact in fixed time with $\hat C_{L,\Nmax} = 2 \sqrt{2}$ with convergence time function in presence of noise
    \begin{align}
    \label{eq:setting:time}
        \hat \Tc(\Ub, N) = \begin{cases}
            t_0 & N \le \frac{L t_0^2}{2} \\
            2 \sqrt{\frac{2N}{L}} + \frac{\frac{2N}{t_0} - \gamma t_0}{\gamma - L} & \text{otherwise}.
        \end{cases}
    \end{align}
    \item \label{item:robutness} $\Diff$ is robust from the beginning.    
    \end{enumerate}
\end{theorem}
\begin{remark}
    Items \ref{item:filt} and \ref{item:robutness} of Theorem \ref{thm:proposed:filtered} relate to the objectives \ref{item:prob:Lipschitz} and \ref{item:prob:robust} from the problem statement. In addition, items \ref{item:errorbound:filt} and \ref{item:settling} of Theorem \ref{thm:proposed:filtered} ensure that objective \ref{item:prob:optimal} is complied for the proposed differentiator. Note that fixed time convergence is guaranteed for $t_0>0$ by Theorem \ref{thm:proposed:filtered}-\ref{item:errorbound:filt}, complying with the goal \ref{item:prob:exact} from the problem statement. The case with $t_0=0$ only ensures finite time convergence, provided here for completeness in the characterization of the proposal.
\end{remark}

\newcommand{\sat}{\operatorname{sat}}
\section{\Rev{Sample-based implementation}}
\label{sec:discreteDiff}

Consider the case that only noisy samples $u_k = f(t_k) + \eta_k$ of the signal $f$ are available at times $t_k = k\Delta$ that are integer multiples of a sampling time $\Delta$.
\Rev{For this case, a sample-based implementation of the proposed differentiator will be constructed as:
\begin{equation}
\label{eq:explicit:sol}
    y_k = \begin{cases}
        0 & \text{if $k < k_0$ or $k = 0$} \\
        y_{\mathrm{s},k_0} & \text{if $k = k_0$ and $k \ge 1$} \\
        y_{k-1} + \sat_{\gamma \Delta}( y_{\mathrm{s},k} - y_{k-1}) & \text{if $k > k_0$},
    \end{cases} 
\end{equation}}
\Rev{with initialization $y_{s,k_0}$ at $k_0 \in \NN_0$, and the saturation function
\begin{equation}
\label{eq:sat}
    \sat_M(x) = \begin{cases}
        x & \text{if $|x| \le M$} \\
        M \sign(x) & \text{otherwise},
    \end{cases}
\end{equation}
and where $y_{\mathrm{s},k}$ denotes the output of the sample-based implementation of the regularized differentiator \eqref{eq:measDiff} to be defined in the following, and $y_{k}$ is the proposed estimate of $\dot f(k\Delta)$. }

\Rev{In can be verified that \eqref{eq:explicit:sol} is the exact solution to:
\begin{equation}
\label{eq:aux:discrete:sys}
    y_{k} = y_{k-1} -  \gamma\Delta \sign(y_{k} - y_{\mathrm{s},k}), \quad y_{k_0} = y_{\mathrm{s},k_0},
\end{equation}
which constitutes the implicit discretization, cf. e.g., \cite{brogliato2021digital}, of the sliding-mode filter \eqref{eq:aux:dyn:sys}. While \eqref{eq:explicit:sol} is preferred from an implementation point of view, \eqref{eq:aux:discrete:sys} will be used in the proofs.}

Since sampled signals (i.e., sequences) cannot lack Lebesgue measurability, the sample-based implementation of the regularized differentiator \eqref{eq:measDiff} coincides with the sample-based implementation of the base differentiator \eqref{eq:proposed:diff}.
Selecting the minimal possible value\footnote{Larger values of $\gamma_k$ are not considered here, because they increase both the complexity of the implementation and the delay of the differentiator without further improving its worst-case accuracy.} for $\gamma_k$ in the results from \cite[Section~7, eq. (52)]{seeber2023}, a corresponding sample-based implementation for $k \ge 1$ is given by
\begin{subequations}
\label{eq:meas:sampled}
    \begin{equation}
        y_{\mathrm{s},k} = \frac{u_k - u_{k - \hat \ell_k}}{\Delta \hat \ell_k},
    \end{equation}
    with the adaptive time difference in samples $\hat \ell_k$ according to
    \begin{equation}
    \label{eq:ellk}
        \hat \ell_k = \begin{cases}
            1 & \hat N_k = 0 \\
            \min\left\{ k, \kmax, \left\lceil \frac{2}{\Delta} \sqrt{\frac{\hat N_k}{L}} \right\rceil \right\} & \hat N_k > 0,
        \end{cases}
    \end{equation}
    \Rev{with window length parameter $\kmax$ (analogous to the continuous-time $\overline{T}$)}, and noise amplitude estimate $\hat N_k$ defined as $\hat N_1 = 0$ and
    \begin{equation}
    \label{eq:Nk}
        \hat N_k = \frac{1}{2} \max_{\substack{\ell \in \{ 2, \ldots, \kmax \} \\ \ell \le k \\ j \in \{1, \ldots, \ell\}}} \left( |Q_{k,\ell,j}| - \frac{L \Delta^2 j (\ell - j)}{2} \right),
    \end{equation}
    for $k \ge 2$, wherein the following abbreviation is used:
    \begin{equation}
    \label{eq:Qk}
        Q_{k,\ell,j} = u_{k - j} - u_k + (u_k - u_{k-\ell}) \frac{j}{\ell}
    \end{equation}

\end{subequations}

To study the accuracy of $y_k$, denote with $\mathcal{D}_\Delta$ the proposed sample-based differentiator, i.e. the operator $[\mathcal{D}_\Delta u](k\Delta)=y_k$. Then, an equivalent definition for worst-case error in the sampled case is borrowed from \cite[Definition 6.1]{seeber2023}:
\begin{definition}
    Let $L,N\in\mathbb{R}_{\geq 0}$ and $\Delta>0$. A sample-based differentiator $\mathcal{D}_\Delta$ is said to have worst-case error $M_N^{L,\Ub}(t)$ from time $t=k\Delta, k\in\mathbb{N}_0$, over the signal class $\mathcal{F}_L^\Ub$ with noise bound $N$ if 
    $$
    M_N^{L,\Ub}(k\Delta)=\sup_{\substack{u=f+\eta \\ \eta\in\mathcal{E}_N \\ f\in\mathcal{F}_L^\Ub}}\sup_{\substack{\ell\in\mathbb{N}_0 \\ l\geq k}}\left|\dot{f}(\ell\Delta)-[\mathcal{D}_\Delta u] (\ell\Delta)\right|.
   $$
\end{definition}

\begin{lemma}
\label{lemma:sampled:meas}
Let $\Delta>0, L >0, N\geq 0$ and consider the sample-based differentiator $\Diff_{\mathrm{m},\Delta}$ with output $y_{\mathrm{s},k} = \Diff_{\mathrm{m},\Delta} u$ defined by \eqref{eq:meas:sampled} with parameter $\kmax \in \mathbb{N} \cup \{ \infty \}$, $\kmax \ge 2$. Let $\overline{N}=L\Delta^2(\kmax-1)^2/2$. Then, for all $N\leq \overline{N}$ and for all $k\in\mathbb{N}$ it holds that:
$$
|y_{\mathrm{s},k}-\dot{f}(k\Delta)|\leq \begin{cases}
2\sqrt{2NL}+\frac{L\Delta}{2} & k\Delta\geq \sqrt{2N/L}\\
L\Delta k+\frac{2N}{\Delta k} & \text{otherwise}.
\end{cases}
$$
\end{lemma}

\begin{theorem}
\label{thm:proposed:filtered:sampled}
Let $\Delta>0, L >0, N\geq 0$ and consider the sample-based differentiator $\Diff_\Delta$ with output $y = \Diff_\Delta u$ defined by \eqref{eq:aux:discrete:sys}, \eqref{eq:meas:sampled} with parameters $\kmax \in \mathbb{N} \cup \{ \infty \}, \kmax \ge 2$, and $\gamma>L$. Let $\overline{N}=L\Delta^2(\kmax-1)^2/2$.
Then, for signals in $\FL$ with noise bound $N < \Nmax$ and all $\Ub \in \mathbb{R}_{\ge 0}^2$, the worst-case error bound
$$
 M_N^{L,\Ub}(k\Delta)\leq 2\sqrt{2NL}+\frac{L\Delta}{2},
$$
holds for all $k\in\mathbb{N}$ with $k\Delta\geq \hat{\mathcal{T}}(\Ub,N)$ wherein
\begin{equation}
\label{eq:setting:time:t0:discrete}
    \hat{\mathcal{T}}(\Ub,N) = 2 \sqrt{\frac{2N}{L}} + \frac{R_1}{\gamma - L} + 3 \Delta \frac{\gamma - \frac{L}{2}}{\gamma - L},
\end{equation}
in case $k_0 = 0$ or in the case of $k_0 > 0$:
\begin{equation}
\label{eq:setting:time:discrete}
    \hat{\mathcal{T}}(\Ub,N) = \begin{cases}
        {k_0}{\Delta} & N \le \frac{L (\Delta k_0)^2}{2} \\
        2 \sqrt{\frac{2N}{L}} + \frac{\frac{2N}{\Delta k_0} - \gamma \Delta k_0}{\gamma - L} + 3 \Delta \frac{\gamma - \frac{L}{2}}{\gamma - L} & \text{otherwise}.
    \end{cases}
\end{equation}
\end{theorem}

\Rev{\begin{remark} \label{Rem:DiscPerf}
    It is worth noting that for $N = 0$, the differentiator satisfies the optimal worst-case bound $M_N^{L,\Ub}(k\Delta)\leq \frac{L\Delta}{2}$ and it is thus quasi-exact as defined in \cite[Definition 6.4]{seeber2023}.
\end{remark}}

\section{Simulation results}
\label{sec:simulations}
For illustration purposes, the proposal in \Revv{\eqref{eq:explicit:sol}} is compared to its non-filtered version in \eqref{eq:meas:sampled} as well as with the sliding-mode based Robust Exact Differentiator (RED). The differentiators are implemented in discrete time with sampling period $\Delta=0.01$, in addition to \Rev{$L=1$ and $\overline{N}=0.08$}. For the RED, we follow the same implementation details as described in \cite{seeber2023} with two parameter sets $\lambda_1=1.5, \lambda_2=1.1$ and $\lambda_1=2.8, \lambda_2=1.96$. \Rev{For the proposal \Revv{\eqref{eq:explicit:sol}}, we selected $\gamma=\lambda_2\geq L$ which, as shown below, results in comparable convergence time as for the RED. Moreover, we consider two cases $k_0=0$ and $k_0=25$ as an arbitrary choice. We limit computational complexity with $\overline{k}=\left\lceil\sqrt{\frac{2\overline{N}}{L\Delta^2}}+1\right\rceil=41$ according to Theorem \ref{thm:proposed:filtered:sampled}.}

\begin{figure*}
\centering
\includegraphics[width=1\textwidth]{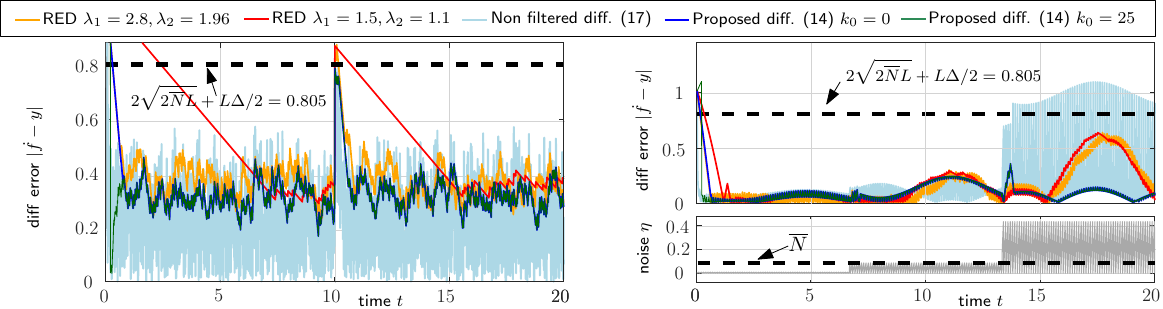}
\vspace{-2em}
\caption{Simulation results with sampling time $\Delta = 0.01$, comparing the RED, the non-filtered differentiator \eqref{eq:meas:sampled} proposed in~\cite{seeber2023}, and the proposed differentiator \Revv{\eqref{eq:explicit:sol}} with $L = 1, N = 0.08$, using $k_0=0$ and $k_0=25$ \Rev{for two different scenarios}. The plots
show the differentiation error $|\dot{f}-y|$ in each case. \Rev{Left: $f(t)=t^2/2+t + 0.5(t-10)\mathds{1}_{t\geq 10}$ and uniformly distributed random noise $\eta(t)$ over $[-N, N]$. Right: $f(t)=(t^2/2+t+\cos(t))/2$ and piece-wise constant noise $\eta(t)$ with different levels of $N$.
}}
\label{fig:sim}
\end{figure*}

\Rev{As a first scenario, consider $f(t)=t^2/2+t + 0.5(t-10)\mathds{1}_{t\geq 10}$, where $\mathds{1}_E$ is the indicator function for the event of $t$ satisfying $E$, introducing a discontinuity in $\dot{f}(t)$ at $t=10$ simulating an impact, as it may occur in a mechanical system, for example. The noise is uniformly distributed in $[-\overline{N}, \overline{N}]$. \Revv{The left part of Fig.~\ref{fig:sim} shows that our approach, given in \Rev{\eqref{eq:explicit:sol}}, attains worst-case accuracy bound guarantees with fixed-time convergence, as the non-filtered differentiator, but with considerably smoother output and similar behavior to the RED, which neither has worst-case accuracy bound guarantees nor fixed-time convergence.} Following from \eqref{eq:setting:time:t0:discrete}, and observing that $R_1=R_2=1$, the convergence times are $\hat{\mathcal{T}}(\Ub,N)=1.8873$ and $\hat{\mathcal{T}}(\Ub,N)=0.5$ for $k_0=0$ and $k_0=25$, respectively. Thus, with $k_0=25$, our proposal improves the convergence time while maintaining the desired accuracy, due to its fixed-time convergence property. Following the impact at $t=10$, all methods re-converge after a transient.}

\Rev{As a second scenario, consider $f(t) = (t^2/2 +t+\cos(t))/2$ with noise $\eta(t) = \nu(t)\text{mod}(t,10\Delta)/(10\Delta)$, where $\nu(t) = \overline{N}(0.1\mathds{1}_{t\leq 20/3} + \mathds{1}_{20/3< t\leq 40/3} + 2\mathds{1}_{40/3< t})$. This piecewise constant noise is used to illustrate the behavior of the differentiators when the noise amplitude is smaller, equal, or larger than $\overline{N}$. \Revv{Note that for $t \in [40/3, 20]$, the assumption that $|\eta(t)| \leq \overline{N}$ is not satisfied as required by our approach. In this case, the optimal worst-case accuracy bounds are not guaranteed anymore by either the filtered or non-filtered differentiators, but a reasonable derivative estimate is still obtained by our proposed differentiator.} }

\section{Conclusions}
\label{sec:conclusions}
A first-order exact differentiator that guarantees optimal worst-case accuracy in a fixed time while being robust to measurement noise and having Lipschitz-continuous output was introduced. The latter combination of properties makes this differentiator unique in comparison with other existing differentiators. The fact that the differentiator is exact means that its output converges to the true derivative in the absence of measurement noise. The maximum Lipschitz constant of the differentiator's output, \Rev{which translates to an increment bound in the sampled case, } is a tunable parameter that allows trading convergence rate for ``smoothness'': the higher this constant, the faster the convergence but the more ``noisy'' the output. Both continuous-time and sample-based versions of the differentiator are developed, the latter being the digitally implementable one. The theoretical guarantees are given for both differentiator versions.

\appendix
\subsection{Auxiliary lemmata}

\begin{lemma}\cite[Adapted from Lemmas 5.7 and 5.8]{seeber2023}
\label{le:bounds:left:diff}
Let $L\geq 0$, suppose that $u\in\mathcal{U}$, and consider a fixed $t>0$ such that $\hat{N}(t)=0$ with $\hat{N}(t)$ defined in \eqref{eq:proposed:diff}. Let $\mu\in(0,t]$ and suppose $Q$ as defined in \eqref{eq:def:aT:t} satisfies
\begin{equation}
\label{eq:Qcondition}
|Q(t,T,\sigma)|\leq \frac{L\sigma(T-\sigma)}{2},
\end{equation}
for all $T\in(0,\mu]$ and all $\sigma\in[0,T]$. Then, the limit
$
\beta := \lim_{T\to0^+}(u(t)-u(t-T))/T
$
exists. Moreover, 
\begin{equation}
\label{eq:difference:sigma}
|u(t-{\sigma})-u(t)+\beta{\sigma}|\leq \frac{L{\sigma}^2}{2},
\end{equation}
holds for all $\sigma\in[0,\mu]$.
\end{lemma}

\begin{lemma}\cite[Lemma 5.9]{seeber2023}
\label{le:difference:sigma}
Let $L, \hat{N}>0, \hat{T}\geq 2\sqrt{\hat{N}/L}, u\in\mathcal{U}$. Let
$
\beta := (u(t)-u(t-\hat{T}))/\hat{T}
$
and suppose 
\begin{equation}
\label{eq:qcondition}
|Q(t,\hat{\sigma},\hat{T})|\leq 2\hat{N}+\frac{L\hat{T}(\hat{\sigma}-\hat{T})}{2}
\end{equation}
is fulfilled for some $\hat{\sigma}\in[\hat{T},t]$. Then, \eqref{eq:difference:sigma} holds for $\sigma=\hat{\sigma}$.
\end{lemma}
\begin{lemma}\cite[Lemma 4.1]{seeber2023}
\label{le:bounds:f}
Let $L\geq 0, f\in\mathcal{F}_L$. Then,
\begin{equation}
\label{eq:bounds:f}
|f(t-\sigma)-f(t)+\dot{f}(t)\sigma|\leq \frac{L\sigma^2}{2}
\end{equation}
holds for all $t\geq 0$ and all $\sigma\in[0,t]$.
\end{lemma}

\subsection{Proofs of all results}

\begin{proof}[\bf Proof of Lemma~\ref{lem:error:boundedness}]
For $t\geq \sqrt{2N/L}$, the result follows from \cite[Theorem 5.1]{seeber2023}. Now, consider $t\in(0,\sqrt{2N/L})$. Note that $\hat{T}(t)=\overline{T}$ leads to the contradiction $t\geq \hat{T}(t)=\overline{T}=\sqrt{2\overline{N}/L}\geq \sqrt{2N/L}$ and for $\hat{T}(t) = t$ the result follows from \cite[Lemma~5.4]{seeber2023}. Distinguish two remaining cases as \textbf{i)} $\hat{T}(t)=0$ and \textbf{ii)} $\hat{T}(t)=2\sqrt{\hat{N}(t)/L}\in(0,t)$. For these cases, we show that \eqref{eq:difference:sigma} holds for $\sigma = t$ with $\beta=\yw(t)$.

Case \textbf{i)} implies $\hat{N}(t)=0$ which in turn implies that \eqref{eq:Qcondition} is fulfilled for all $T\in(0,\mu], \sigma\in[0,T]$ and $\mu=\min\{t,\overline{T}\}=t$ due to \eqref{eq:def:Nhat:t}.
Applying Lemma \ref{le:bounds:left:diff} with $\mu = t$, it follows that $\yw(t) = \beta$ and that \eqref{eq:difference:sigma} holds for ${\sigma}=t$.

In case \textbf{ii)}, $\hat{T}(t) \in (0,t)$ and \eqref{eq:qcondition} is fulfilled for $\hat{\sigma} = t$, $\hat N = \hat N(t)$, $\hat T = \hat T(t)$ due to the construction of $\hat{N}(t)$ in \eqref{eq:def:Nhat:t} (see \cite[Remark 5.10]{seeber2023}).
Thus, applying Lemma \ref{le:difference:sigma} yields that \eqref{eq:difference:sigma} holds for ${\sigma} = t$ with $\beta=\yw(t)$.

In both cases,
$
|f(0)-f(t)+\yw(t) t|\leq \frac{Lt^2}{2}+2N
$
is obtained due to $|\eta(t)|\leq N$. Combine this inequality with \eqref{eq:bounds:f} from Lemma \ref{le:bounds:f}, setting $\sigma = t$, to obtain
$
|\yw(t)t-\dot{f}(t)t|\leq Lt^2 + 2N.
$
Division by $t$ proves the claim.
\end{proof}

\begin{proof}[\bf  Proof of Lemma \ref{lemma:sampled:meas}]
The case $k\Delta\geq \sqrt{2N/L}$ follows from \cite[Theorem 7.1]{seeber2023}. In the other case, $\hat{\ell}_k = k$ allows to apply \cite[Lemma~5.4]{seeber2023} with $\hat T = \Delta \hat \ell_k$, and $\hat{\ell}_k=\overline{k}$ implies the contradiction %
$\overline{k}\Delta \le k\Delta < \sqrt{2N/L} \le \sqrt{2\overline{N}/L} = (\overline{k}-1)\Delta$.
The remaining case is $\hat{\ell}_k=\left\lceil \frac{2}{\Delta} \sqrt{\frac{\hat N_k}{L}} \right\rceil\in[1,k)$. In that case, apply Lemma~\ref{le:difference:sigma} with $t = \hat\sigma = k\Delta$, $\hat T = \hat\ell_k \Delta$ while noting that $|Q(k\Delta,k\Delta,\hat\ell_k\Delta)| = |Q_{k,k,\hat\ell_k}| \le 2 \hat N_k + L\Delta^2\hat\ell_k(k-\hat\ell_k)/2$ is fulfilled due to \eqref{eq:Qk}, \eqref{eq:def:aT:t}, and \eqref{eq:Nk}.
That lemma yields $|u_0 - u_k + \beta k \Delta| \le L k^2 \Delta^2/2$ with $\beta = y_{\mathrm{s},k}$, implying $|f_0 - f_k + y_{\mathrm{s},k} k \Delta| \le L k^2 \Delta^2/2 + 2 N$ due to $|\eta_k| \le N$ and yielding $|y_{\mathrm{s},k} k\Delta - \dot f(k\Delta) k \Delta| \le L k^2 \Delta^2 + 2 N$ after combining with the statement of Lemma~\ref{le:bounds:f}.
Finally, divide by $k\Delta$.
\end{proof}

\begin{proof}[\bf Proof of Proposition~\ref{prop:proposed:exact}]
For item~\ref{item:meas}, note that every $u \in \mathcal{U}$ is locally bounded on $\RR_{\ge 0}$.
Lemma~\ref{lem:error:boundedness} then implies that $\yw = \Diffw u$ is locally bounded on $\RR_{> 0}$, which allows to conclude Lebesgue measurability of $\Diffm u = \yw^{\ddag}$ using \cite[Lemma 2]{aldana2023exact}.

Regarding item \ref{item:errorbound}, Lemma \ref{lem:error:boundedness} ensures that the output $\yw = \Diffw (f+\eta)$ of $\Diffw$ satisfies \eqref{eq:boundsyw} 
for all $f \in \FL$, $\eta \in \EN$.
Note that $\limsup_{\varepsilon\to 0^+}\dot{f}(\tau-\varepsilon) = \dot{f}(\tau)$ since $\dot{f}$ is Lipschitz-continuous.
Therefore, it follows that 
\begin{equation}
\begin{aligned}
    &\big|\dot f(\tau) - \limsup_{\varepsilon\to 0^+}\yw(\tau-\varepsilon)\big| \\&= 
    \left|\limsup_{\varepsilon\to 0^+}\left(\dot{f}(\tau-\varepsilon) - \yw(\tau-\varepsilon)\right) \right| \\
    &\le \begin{cases}
        Lt + \frac{2N}{t} & \text{if }t\in(0,\sqrt{2N/L}) \\
        2\sqrt{2NL} & t\geq \sqrt{2N/L},
        \end{cases}
\end{aligned}
\end{equation}

for $\tau\geq t$ since $L\tau + \frac{2N}{\tau}$ is decreasing for $\tau\in(0,\sqrt{2N / L})$. The same conclusion applies to $\liminf_{\varepsilon\to 0^+}\yw(t-\varepsilon)$ and in turn to $\ym(t)$, completing the proof for this item.

For item \ref{item:robustDw}, note that $Q_N^{L,\Ub}(t) \le M_N^{L,\Ub}(t) + M_0^{L,\Ub}(t)$.
Robustness almost from the beginning follows from the fact that $\lim_{N \to 0^+} M_N^{L,\Ub}(t) = 0$ for all $t > 0$ due to item~\ref{item:errorbound}.
\end{proof}

\begin{proof}[\bf Proof of Theorem~\ref{thm:proposed:filtered}]
Item \ref{item:filt} follows by noting that $\dot{y}(t)$ exists almost everywhere and is bounded by $\gamma$. Therefore, $y(t)$ is Lipschitz. 

Item \ref{item:errorbound:filt} is already proven in \cite[Theorem 1, item b)]{aldana2023exact}; because of its relevance for item \ref{item:settling}, we recall the proof for completeness. Let $f \in \FL^\Ub$, $\eta \in \EN$ with $N < \Nmax$, and define the differentiation error $e(t) = y(t) - \dot f(t)$ with $y = \Diff (f + \eta)$.
From \eqref{eq:aux:dyn:sys}, $e$ satisfies
\begin{equation}
\label{eq:e2sys}
\dot{e}(t) =  - \gamma\sign( e(t) - \etam(t) ) -\ddot{f}(t),
\end{equation}
and $|e(0)| = |y(0)-\dot{f}(0)| = |\dot{f}(0)| \le R_1$, with $\etam = \ym - \dot f = \Diffm f - \dot f$ being Lebesgue measurable according to Proposition \ref{prop:proposed:exact}-\ref{item:meas} and bounded by
\begin{equation}
\label{eq:boundnoise}
    |\etam(t)| \le 2 \sqrt{2 N L},
\end{equation}
for all $t > \tilde \Tc(N) := \sqrt{2 N/L}$ according to Proposition~\ref{prop:proposed:exact}-\ref{item:errorbound}.
For $t \in [0, \tilde \Tc(N)]$ it follows from $|\dot e| \le \gamma + L$ that
\begin{equation}
\label{eq:bounde}
    |e(t)| \le (\gamma+L)t + |e(0)| \le (\gamma+L) \tilde \Tc(N) + R_1
\end{equation}
holds.
For $t > \tilde \Tc(N)$, consider $V(e) = |e|$ as a Lyapunov function.
Then, $V = |e| > 2 \sqrt{2 N L}$ implies that its time derivative $\dot V$ along \eqref{eq:e2sys} satisfies
\begin{equation}
\label{eq:Vdot}
    \dot V = - \gamma \sign(e) \sign(e - \etam(t)) - \ddot f(t) \sign(e) \le -\gamma + L,
\end{equation}
since $\sign(e-\etam) = \sign(e)$ due to \eqref{eq:boundnoise}.
Noting that $V(e(\tilde \Tc(N))) \le (\gamma + L) \tilde \Tc(N) + R_1$, it will now be shown using the comparison principle that $V(e(t)) \le 2 \sqrt{2 N L}$ holds for all $t \ge \hat \Tc(\Ub,N)$,
proving the claim that the worst-case error satisfies $M_L^{\Ub,N}(\hat \Tc(\Ub,N)) \le 2 \sqrt{2 N L} = \hat C_{L,\Nmax} \sqrt{NL}$ with $\hat C_{L,\Nmax} = 2\sqrt{2}$.
To see this, suppose to the contrary that $V(e(t)) > 2\sqrt{2 NL}$ holds for some $t \ge \hat\Tc(\Ub,N)$.
Then, the differential inequality \eqref{eq:Vdot} may be integrated backward in time to obtain $V(e(t)) > 2 \sqrt{2 NL}$ for all $t \in [\tilde \Tc(N), \hat \Tc(\Ub,N)]$ along with the contradiction and using $2 \sqrt{2 N L} = 2 L \tilde \Tc(N)$:
\begin{align}
&V(e(\tilde \Tc(N))) \nonumber\\
&\ge V(e(\hat \Tc(\Ub,N))) + (\gamma - L) (\hat \Tc(\Ub,N) - \tilde \Tc(N)) \nonumber \\
&= (\gamma + L) \tilde \Tc(N) + R_1.
\end{align}

For item \ref{item:settling}, if $N\leq L t_0^2/2$, then $|\etam(t)|\leq 2\sqrt{2NL}$ for all $t\geq t_0\geq \tilde \Tc(N)$ due to Proposition~\ref{prop:proposed:exact}-\ref{item:errorbound}. Hence, $|e(t_0)|=|y(t_0)-\dot{f}(t_0)|=|\ym(t_0)-\dot{f}(t_0)|\leq 2\sqrt{2NL}$. Using the same reasoning as in item b), $|e(t)|\leq 2\sqrt{2NL}$ remains invariant for all $t\geq t_0=\hat \Tc(\Ub,N)$. 

Now, consider the case with $N\geq L t_0^2/2>0$. Then, 
$$
\begin{aligned}
|e(t_0)| =|\etam(t_0)| \leq \frac{2N}{t_0} + Lt_0
\end{aligned}
$$
due to Lemma \ref{lem:error:boundedness}. Similarly, as in the proof of item b), for $t\in[t_0,\tilde \Tc(N)]$ it follows from $|\dot{e}|\leq L+\gamma$ that
$$
\begin{aligned}
|e(t)|&\leq (\gamma+L)(t-t_0) + |e(t_0)|
\\&\leq (\gamma+L)(t-t_0) + \frac{2N}{t_0} + Lt_0 \\
&\leq (\gamma + L)t + \frac{2N}{t_0}-\gamma t_0.
\end{aligned}
$$
Henceforth, \eqref{eq:bounde} follows for all $t\in[t_0,\tilde \Tc(N)]$ replacing $R$ by $\frac{2N}{t_0}-\gamma t_0$. Therefore, $V(\tilde \Tc(N))\leq (\gamma+L)\tilde \Tc(N) + \left(\frac{2N}{t_0}-\gamma t_0\right)$. The proof follows in the same way as the rest of the proof for item \ref{item:errorbound:filt}, with such replacement for $R_1$.

Finally, we show item \ref{item:robutness}. Let $f \in \FL^\Ub$, $\eta \in \EN$ for $N \le \Nmax$ and denote with $y_0(t)=[\Diffm f](t)$ the output of \eqref{eq:aux:dyn:sys} with $N=0$, which complies $y_0(t)=y(t)$ for $t\in[0,t_0]$ with $t_0\geq 0$. Moreover, according to item~\ref{item:errorbound:filt}, $y_0$ fulfills $y_0(t) = \dot f(t)$ for all $t\geq \hat \Tc(\Ub,0)$. For $t > \hat \Tc(\Ub,N) \ge \hat \Tc(\Ub,0)$, it follows that
\begin{equation}
    |y_0(t) - y(t)| = |y(t) - \dot f(t)| \le M_N^{L,\Ub}(t) \le 2 \sqrt{2 N L}.
\end{equation}
For $t \in [t_0, \max(\tilde \Tc(N),t_0)]$,
\begin{equation}
    |y_0(t) - y(t)| \le 2 \gamma (t-t_0) \le 2 \gamma \max(\sqrt{2N/L}-t_0,0)
\end{equation}
holds by nature of \eqref{eq:aux:dyn:sys}.
For $t \in (\max(\tilde \Tc(N),t_0),\hat \Tc(\Ub,N)]$, finally, note that $y_0$ satisfies $\dot y_0 = -\gamma \sign(y_0 - \dot f)$, because $\Diffm f = \dot f$, and that \eqref{eq:aux:dyn:sys} on that interval is a perturbed version of that system with the perturbation of initial condition and right-hand side bounded according to
\begin{subequations}
\begin{align}
    |y_0(\sqrt{2N/L}) - y(\sqrt{2N/L})| &\le 2 \gamma  \max(\sqrt{2N/L}-t_0,0), \\
    |\dot f(t) - \ym(t)| &\le 2 \sqrt{2 N L}.
\end{align}
\end{subequations}
Hence, it follows from continuous dependence of solutions \cite[Theorem 1, page 87]{Filippov1988} that for every $\epsilon>0$ there exists an $N > 0$ such that
\begin{equation}
    |y_0(t) - y(t)| \le \epsilon,
\end{equation}
holds for all $t \in [ \max(\sqrt{2N/L}-t_0,0),\hat \Tc(\Ub,N)]$.
As a consequence of the three considered cases,
\begin{equation}
    \limsup_{N \to 0^+} \sup_{\tau \ge 0} |y_0(\tau) - y(\tau)| = 0
\end{equation}
holds, completing the proof. 
\end{proof}

\begin{proof}[\bf Proof of Theorem~\ref{thm:proposed:filtered:sampled}]
For the case $k_0 = 0$, let $f \in \FL^\Ub$, $\eta \in \EN$ with $N \leq \Nmax$, and define the differentiation error $e_k = y_k - \dot f(k\Delta)$ with $y_k = \Diff_\Delta (f(k\Delta) + \eta_k)$ and $\eta_k=\eta(k\Delta)$.
From \eqref{eq:aux:discrete:sys}, $e_k$ satisfies
\begin{equation}
\label{eq:e2sys:discrete}
e_k = e_{k-1} - \gamma\Delta\sign( e_k - \eta_{\mathrm{m},k} ) +g_k,
\end{equation}
with $g_k:=\dot{f}((k-1)\Delta)-\dot{f}(k\Delta)$ and $\eta_{\mathrm{m},k} = y_{\mathrm{s},k} - \dot f(k\Delta) = [\Diff_{\mathrm{m},\Delta} f - \dot f](k\Delta)$.  Note that $|g_k|\leq \Delta L$ and thus,
$
|e_k-e_{k-1}|\leq \Delta(\gamma+L)
$
which, given $|e_0|=|\dot{f}(0)|\leq R_1$, results in 
\begin{equation}
\label{eq:sampled:R}
|e_k|\leq R_1+k\Delta(\gamma+L),
\end{equation}
for all $k\geq 0$. Lemma \ref{lemma:sampled:meas} implies that $|\eta_{\mathrm{m},k}|\leq 2\sqrt{2NL}+{L\Delta}/{2}$ for all $k\Delta\geq \tilde{\mathcal{T}}(N)=\sqrt{2N/L}$. Now, set $V(e_k)=|e_k|$ as a Lyapunov function candidate to obtain:
\begin{equation}
\label{eq:discrete:lyap1}
V(e_k)-V(e_{k-1}) = |e_k|-|e_k +\gamma\Delta\sign( e_k - \eta_{\mathrm{m},k} ) -g_k|.
\end{equation}
Consider $k\Delta\geq \tilde{\mathcal{T}}(N)$ and $|e_k| > 2\sqrt{2NL}+{L\Delta}/{2}$. Then, \eqref{eq:discrete:lyap1} implies:
\begin{equation}
\label{eq:discrete:lyap2}
\begin{aligned}
V(e_k)-V(e_{k-1})&\leq |e_k|-|e_k+\gamma\Delta\sign(e_k)-g_k| \\
&= |e_k|-|    |e_k|+\gamma\Delta-g_k\sign(e_k)|\\
&\leq -\Delta(\gamma-L).
\end{aligned}
\end{equation}
Set $k_1=\lceil\tilde{\mathcal{T}}(N)/\Delta\rceil$ such that $V(e_{k_1})\leq R_1+k_1\Delta(\gamma+L)$ and assume $V(e_{k_2}) > 2 \sqrt{2NL} + L\Delta/2$ with $k_2 := \lceil\hat{\mathcal{T}}(\Ub,N)/\Delta\rceil$. Then, noting that $k_1 \Delta \le \tilde{\mathcal{T}}(N) + \Delta$, the contradiction
\begin{align}
    V(e_{k_1}) &\ge V(e_{k_2}) + (k_2 - k_1) \Delta(\gamma - L) \nonumber \\
    &> 2 \sqrt{2NL} + \frac{L\Delta}{2} + (\hat{\mathcal{T}}(\Ub,N) - \tilde{\mathcal{T}}(N) - 2 \Delta)(\gamma -L) \nonumber \\
    &\ge R_1 + k_1 \Delta (\gamma + L),
\end{align}
proves that $|e_{k_2}| \le 2 \sqrt{2NL} + L \Delta/2$ holds.
Now, given $|e_{k-1}|\leq 2\sqrt{2NL}+L\Delta/2$, assume $|e_k|\geq 2\sqrt{2NL}+L\Delta/2$. Thus, similarly as before:
$$
|e_{k-1}|=|e_k+\gamma\Delta\text{sign}(e_k)-g_k|\geq |e_k|\geq 2\sqrt{2NL}+L\Delta/2,
$$
which is a contradiction, and thus, $|e_k|\leq 2\sqrt{2NL}+L\Delta/2$ is maintained invariant for all $k\geq k_2$.

For the case $k_0 > 0$, if $N\leq L (k_0\Delta)^2/2$, then $|\eta_{\mathrm{m},k}|\leq 2\sqrt{2NL}+L\Delta/2$ for all $k\geq k_0\geq \tilde \Tc(N)/\Delta$ due to Lemma~\ref{lemma:sampled:meas}. Hence, $|e_{k_0}|=|y_{k_0}-\dot{f}(k_0\Delta)|=|y_{\mathrm{m},k_0}-\dot{f}(t_0\Delta)|\leq 2\sqrt{2NL}+L\Delta/2$. Using the same reasoning as in the case $k_0 = 0$, $|e_k|\leq 2\sqrt{2NL}+L\Delta/2$ remains invariant for all $k\Delta\geq k_0=\hat \Tc(\Ub,N)$. Now, consider the case with $N\geq L (\Delta k_0)^2/2>0$. Then, 
$$
\begin{aligned}
|e_{k_0}| =|\eta_{\mathrm{m},k_0}| \leq \frac{2N}{\Delta k_0} + \Delta Lk_0,
\end{aligned}
$$
due to Lemma \ref{lemma:sampled:meas}. Similarly as in the proof for the case $k_0 = 0$, for $k\in[k_0,\tilde \Tc(N)/\Delta]$ it follows from $|e_k-e_{k-1}|\leq \Delta(L+\gamma)$ that
$$
\begin{aligned}
|e_k|&\leq \Delta(\gamma+L)(k-k_0) + |e_{k_0}|
\\&\leq \Delta(\gamma+L)(k-k_0) + \frac{2N}{\Delta k_0} + \Delta Lk_0 \\
&\leq \Delta(\gamma + L)k + \frac{2N}{\Delta k_0}-\gamma\Delta k_0.
\end{aligned}
$$
Henceforth, \eqref{eq:sampled:R} follows for all $k\in[k_0,\tilde \Tc(N)/\Delta]$ replacing $R_1$ by $\frac{2N}{\Delta k_0}-\gamma\Delta k_0$. The proof follows in the same way as the rest of the proof for the case $k_0 = 0$, with such $R_1$.
\end{proof}

\end{document}